\documentstyle[12pt,psfig,aasms4]{article}

\def\msun{$M_{\odot}$}

\def\etal{{\it et al. }}

\def\asec{\ifmmode ^{\prime\prime}\else$^{\prime\prime}$\fi}
\def\amin{\ifmmode ^{\prime}\else$^{\prime}$\fi}
\def\degs{\ifmmode ^{\circ}\else$^{\circ}$\fi}
\newbox\grsign \setbox\grsign=\hbox{$>$}
\newdimen\grdimen \grdimen=\ht\grsign
\newbox\laxbox \newbox\gaxbox
\setbox\gaxbox=\hbox{\raise.5ex\hbox{$>$}\llap
     {\lower.5ex\hbox{$\sim$}}}\ht1=\grdimen\dp1=0pt
\setbox\laxbox=\hbox{\raise.5ex\hbox{$<$}\llap
     {\lower.5ex\hbox{$\sim$}}}\ht2=\grdimen\dp2=0pt

\def\msun{$M_{\odot}$}

\received{July 5, 1996}

\begin{document}

\title{RXTE observations of GRS 1915+105}

\author{J. Greiner}
\affil{Max-Planck-Institute for extraterrestrial Physics, 85740 Garching, 
Germany}
\author{E.H. Morgan, R.A. Remillard}
\affil{Center for Space Research, MIT, Cambridge, MA 02139, USA}

\authoremail{jcg@mpe-garching.mpg.de}

\begin{abstract}
The galactic superluminal motion source GRS 1915+105 was observed with the
RXTE satellite at several occasions during its ongoing active state.
The observed X-ray intensity changes drastically on a variety of time scales
ranging from sub-seconds to days. 
In particular, the source exhibits quasi-periodic brightness sputters with 
varying duration and repetition time scale. These episodes occur 
occasionally, while the more common 
X-ray intensity variations are faster with much smaller amplitudes. 
The spectrum during the brightness sputters is remarkably different from
the spectrum of the mean high state emission.
We argue that such sputtering episodes are
possibly caused by a major accretion disk instability. 
Based on the coincidence in time of two radio flares following
the observed X-ray sputtering episodes
we speculate that superluminal ejections (as observed from GRS 1915+105
during earlier activity periods) are related to episodes of large 
amplitude X-ray variations.
\end{abstract}


\section{Introduction}

On 1992 August 15 the WATCH detectors on {\it Granat} discovered a new X-ray
transient (Castro-Tirado \etal 1992) which was designated GRS 1915+105.
Follow-up monitoring with WATCH and BATSE (Harmon \etal 1992) revealed an
unusual light curve consisting of a 3 month rise, an 8 month high-intensity
plateau phase, and a 2 month decline.
A variable radio source was found with the VLA (Mirabel \etal 1993a)
inside the $\pm$10\asec\ X-ray error circle (Greiner 1993), as well as 
a variable infrared source (Mirabel \etal 1993b, Castro-Tirado \etal 1993).
The soft X-ray spectrum as seen with the ROSAT PSPC on several occasions
is strongly absorbed 
(N$_{\rm H}\approx 5\times10^{22}$ cm$^{-2}$; Greiner 1993) supporting the 
earlier conjecture that GRS 1915+105 is located behind the Sagittarius arm, 
i.e. at a distance greater than 8 kpc (Greiner \etal 1994). Indeed, later HI
measurements suggest a distance of about 12.5 kpc (Mirabel \& Rodriguez 1994).
The optical counterpart (at the radio position)  was only detected in the 
I band at 23.4 mag, while it is fainter than 25.9, 26.1 and 26.1 mag in B, V, 
and R, respectively (Bo\"er, Greiner \& Motch 1996).

An outburst with a light curve similar to the one described above
(but with shorter duration) occurred between December 1993 and April 1994,
at which time GRS~1915+105 was observed in a remarkably high radio state. 
Further radio monitoring revealed radio structures travelling at apparently 
superluminal speed (Mirabel \& Rodriguez 1994) making GRS 1915+105 the first 
superluminal source in the Galaxy. Until then, apparent superluminal motion 
was only observed in AGN, the central engines 
of which are generally believed to be massive black holes. This similarity 
suggests that GRS 1915+105 harbors a stellar-sized black hole. Indirect 
support for this hypothesis was provided by the discovery of the second galactic
superluminal motion source GRO J1655--40 (Zhang \etal 1994), for which a mass
function larger than 3 \msun\, was subsequently established by optical
radial velocity measurements (Bailyn \etal 1995).

\section{ASM discovery of renewed activity in GRS 1915+105}

The RXTE All Sky Monitor (ASM) observes bright X-ray sources in the range of 
$\sim$2 to 12 keV using position-sensitive proportional counters placed below 
coded masks.  X-ray intensity measurements are derived from the deconvolution 
of overlapping mask shadows from each X-ray source in a camera's field of view. 
There are 3 cameras (SSCs) in the ASM, and the field of view of each camera is 
6$\times$90 degrees FWHM.  Histograms of detected
counts vs. anode wire position are telemetered for each SSC in 3 separate
energy bands (1.3--3.0 keV, 3.0--4.8 keV, and 4.8--12.2 kev).  Detector problems
were experienced soon after launch, but performance was stabilized with an
operating duty cycle $\sim$40\% (limited to low-background regions of the XTE 
orbit). 
This data rate provides $\sim$7 celestial scans per day, with sky coverage
obtained in a series of 90 s stationary exposures (``dwells") followed by an
instrument rotation of 6 degrees. Data collection has been ongoing with at
least 1 SSC since 1996 February 21.

 The data analysis utilizes all 8 anode wires in SSC1 and 6 anode
wires in each of SSC2 and SSC3.  The fine details of the instrument
calibration are still being refined, and the present analysis is subject
to a systematic error of about 3\% for uncrowded source regions
(such as around GRS 1915+105),
as inferred from the rms deviations observed in the Crab Nebula.
Further details of the instrument and the data analysis methods are
provided by Levine \etal (1996).  For the ASM observations of GRS 1915+105
reported below, we select results from individual SSC dwells and normalize
all of the results to the throughput of SSC1, in which the Crab yields a total
counts rate of 73 SSC c/s.  
For the present investigation, we exclude measurements in which
the angle between the X-ray source and the Earth is less than
75\degs.  We also exclude observations in which a camera's
field of view contains more than 15 potential X-ray sources
or observations in which any 2 sources are within a rectangle
of 0\fdg1\,$\times$\,6\fdg0 aligned perpendicular to the anode wires.

The ASM light curve for the period 1996 February 21 to June 21
is shown in Fig. \ref{lcasm}.  
Extreme and evolving levels of X-ray variability were confirmed by the 
appearance of rapid variations (s to min) seen in the ASM time series data
(collected in parallel with position histogram data) during many dwells
when GRS 1915+105 was in an SSC field of view.  An ASM alert 
led to RXTE pointed observations with the PCA and HEXTE instruments. The 
times of the pointed observations are marked with an asterisk in the 
top panel of Fig. \ref{lcasm}.

\section{Pointed RXTE Observations}

The Proportional Counter Array (PCA) on RXTE (Zhang \etal 1993) consists 
of 5 nearly identical
xenon-filled proportional counter units (PCUs) each with 1250 cm$^2$ of
collecting area. The detectors are sensitive to X-rays between 2 and
60 keV with 17\% energy resolution at 6 keV.  The Crab produces
2500 cts/s/PCU (2--60 keV) while the background count rate is typically 
$\sim$20 cts/s/PCU. On April 6, 9, 17 and May 14, 
only 3 PCUs were used due to an operational problem. On
April 15 the gain of all 5 PCUs was lowered by 30\%.
Due to the steep energy spectrum (see below) we discuss the PCA but not the 
HEXTE data in this short communication.

\subsection{Temporal characteristics}

The X-ray light curves reveal
a variety of features, one of which is large amplitude intensity variations.
This is very similar to the behavior seen from dwell to dwell in the ASM,
as well as in the light curves of individual dwells.

We identify the following properties in the light curves of GRS 1915+105
(see Figs. \ref{lcall}, \ref{Apr06lull} and \ref{May26lull}):
\begin{itemize}
\item Repeating pattern of brightness sputters: In 5 of the pointed PCA 
observations,
 Apr. 6, May 21 and 26 and Jun 16 and 19  we find large, eclipse-like dips in 
the X-ray flux, which we call sputters. During these sputters the flux 
drops from $\approx$2--3 Crab to a momentary lull at 
about 100--500  mCrab and
then shoots up again. The spectrum softens dramatically during the 
brightness sputters. 
The sputters occur quasiperiodically with a recurrence time of about 250 sec 
during the first observation (see top part of Fig. \ref{Apr06lull}). 
The decline to the low level (lull)  typically takes 3--5 sec and the rise 
to the high level 5--10 sec. The lull duration varies irregularly,
with a mean value of between 30--50 sec during the first observation 
and only a few seconds in later observations.
The low-intensity value during the lull is surprisingly constant with
time during the first observations. It is also interesting to note that the 
rms variability during the brightness sputters is much smaller than elsewhere.
During the June 16 and 19 observations the brightness sputters all have a 
small flare lasting ~10s at the start of the sputter.
\item Large-amplitude oscillations: On Apr. 6 the X-ray intensity
starts oscillating with increasing amplitude towards the lull, with  
the last oscillation before the lull nearly reaching the low-intensity level 
of the lull (top of Fig. \ref{Apr06lull}). 
On May 26 we see extremely large amplitude oscillations with an
amplitude of nearly 3 Crab and periods of 30--100s. 
\item Flare following lull: On one occasion on Apr. 6 the intensity remained 
low for over 300 sec and there is a large (to $>$3 Crab) flare following 
this lull, that is spectrally invariable, in sharp contrast to the spectral 
softening seen during the brightness sputters. After the flare there is a  
long delay before sputtering resumes (Fig. \ref{Apr06lull}). 
\item Brief flares in prolonged lulls: During a major lull observed on May 26 
(see Fig. \ref{May26lull}) the source exhibited a flare lasting 5s. Before this
flare the intensity went below the QPO floor. About
400 seconds later another sharp flare occurs as the source intensity climbs 
back up. This complex pattern is repeated almost identically 4925 seconds 
later (Fig. \ref{May26lull}).
Unfortunately, the observation ended before we could see if the pattern repeated
more than twice. 
\item Fast oscillations: Between the episodes of large-amplitude variations
the variations are more regular, developing into
clearly visible quasi-periodic oscillations (20 mHz- 67 Hz) which have rich
phenomenology including spectral characteristics and dynamically changing
amplitudes and frequencies. These features will be discussed in a
following paper (Morgan \etal 1996).
\end{itemize}

The ASM light curve of GRS 1915+105 is marked by an interval of extreme
and random variations that were in progress on Feb. 21 and continued until
April 7 (MJD 50180; see Fig. \ref{lcasm}).  A later period that can be 
characterized as a series of X-ray flares (hours to days) occurred from 
May 17 to July 16 (MJD 50220-20280), and this interval was
followed by 25 days of apparently steady emission at 0.51 Crab.  The PCA
observations with highest amplitudes of both X-ray flux and variability
occurred on April 6, May 20--26, and June 11--19, and these sample both 
the intervals of random variability and X-ray flaring seen with the ASM.
A more detailed analysis of the temporal evolution of the quasi-periodic
oscillations and their energy dependence are presented in a
forthcoming paper (Morgan \etal 1996).

\subsection{Spectral characteristics}

While a complete spectral analysis is beyond the scope of this letter,
a few general features can be deduced already at this early stage.
The spectra during the April 9 through 29 observations are complex and
rapidly variable. Most of these spectra are rather steep and fall off
beyond 20 keV.

The hardness ratio panels in Figs. \ref{Apr06lull} and \ref{May26lull} already
demonstrate that the spectrum of the X-ray emission during the  lulls
is softer than during the high-intensity states, i.e. these lulls are 
not caused by absorption or any low-energy cut-off.
We have selected photons (for individual layers and single PCA units using
the ftools command saextrct)
at different time intervals corresponding to these two intensity states.
Slew data at the end of the observation was
used for background subtraction and the spectral fitting was done using the
XSPEC package (Shafer \etal 1991).
The gross energy distribution of the high-intensity emission (on April 6) can 
be described by a power law model with photon index $\alpha = -1.6$ and an
high-energy cutoff around 5 keV while the high-intensity spectra on
May 20 and 26 require a second component above $\approx$15 keV.
The spectrum during the lulls (April 6, May 26) is well represented
by a two component model, e.g. a disk blackbody plus a power law with photon 
index $\alpha = -2.2$ without a high-energy cutoff. 
The spectrum during the rebound is an extremely steep pure power law 
($\alpha \approx -4.5$).
As can be inferred from the hardness ratio plot, there are no major
spectral changes during the decay phase between lulls before the onset of
the large-amplitude oscillations. But during these oscillations the spectrum
gets harder and softer than the mean on the same time scale as the oscillations.

\section{Discussion}

The  previously known  properties of GRS 1915+105 had signified an unusual
nature, compared to the known types of high-energy transients:
The relativistic radio jets, slow rise time, the spectral characteristics, 
as well as the lack of a systematic flux decay are markedly different from 
black hole transients like e.g. GRS 1124--68 or GRO J0422+32.
With our data from RXTE observations of GRS 1915+105 these 
differences are even more pronounced.

The absence of pulsed emission (Morgan \etal 1996)
suggests that the observed X-ray emission
does not come directly from the surface of a rotating, magnetic compact object. 
On the other hand, the strong
variability during the high-intensity states on time scales well below one 
second indicates that the emission region cannot be a large and hot corona. 
We therefore tentatively assign this emission to an accretion disk.

GRS 1915+105 as observed in April 1996 with RXTE 
is at least a factor of 3 more intense than it was when observed
by ROSAT (Greiner \etal 1994) and ASCA 
(Nagase \etal 1994, Ebisawa \& White 1995) during  previous active states.
At a distance of GRS 1915+105 of 12.5 kpc (Mirabel \& Rodriguez 1994),
the mean, unabsorbed luminosity in the 1--25 keV range is of the order of 
10$^{39}$ erg/s during the high-intensity states (a factor three more than
has been noted previously, e.g. Sazonov \& Sunyaev 1996). This is well
above the Eddington luminosity for a neutron star with any reasonable  mass,
suggesting that the system contains a black hole.

As evidenced by the energy spectra (softening), the brightness sputters are not
absorption dips. Occultation events are unlikely due to two reasons: 
First, the occulting body would have to be large in comparison to the projected
X-ray emitting area (in order to explain the short rise and fall times)
which is difficult to reconcile if the emitting area is an accretion disk,
even if only a fraction of the disk dominates the emission.
Second, unless such an occulter were related to the large-amplitude 
oscillations in the X-ray flux, there is no reason to have the intensity drops
always occur during times following large amplitude oscillations.
It is therefore more plausible to relate the large-amplitude
oscillations as well as the sputters to a single phenomenon, e.g. an 
accretion disk instability. This is supported by the fact that the
large-amplitude oscillations show similar spectral changes as the sputters.

If one relates the strongly variable X-ray emission to the inner part of an 
accretion disk, and since the spectrum during the lulls is markedly different,
consequently one has to suppose that the inner part of the accretion disk
is changed drastically.
The accretion flow in the inner regions of an accretion disk around a
compact object has been proposed to undergo a sonic transition before
falling supersonically onto the compact object (Matsumoto \etal 1985).
As the inner accretion disk is re-filled
rather quickly, it is certainly always far from being in
a steady state. In fact, the observed X-ray spectrum in the 2--20 keV range 
is completely dissimilar to a steady, multicolor disk blackbody
except during the lulls.

It is worth noting that the rebound after lulls does not correspond to the
release of energy which is ``missing" over the lull duration. 
This is a suggestive hint
that the energy is channeled into a different energy form, possibly kinetic.
Episodes of large amplitude variations
(= major instabilities in our interpretation) have been observed
on April 6, May 20 and 26 as well as June 16 and 19, 1996 (Fig. \ref{lcall}).
Interestingly, these times coincide with radio flares 
(with a delay between 1--5 days) reported for the 
period of May 23 through 26 (Pooley 1996), and another one 
during June 17 through 22 (Pooley, priv. comm.). 
Therefore, the episodes of large amplitude X-ray variations may be related
to the formation of jets.
We also note that during these episodes the X-ray spectrum shows
a distinct hard component (though varying)  which seems to be missing at other 
times. This component might enable the detection of GRS 1915+105
by BATSE (see Zhang \etal 1996 for the May 20 through 26 period).
Previous radio monitoring of GRS 1915+105 and GRO J1655--40 has resulted in 
the perplexing result that not all X-ray active states were accompanied by 
radio emission (e.g. the April 1995 and 
the August 1995 activity period of GRO J1655--40, Foster \etal 1996),
suggesting that radio emission (jets) are not related simply to
X-ray high states. If the correlation of radio emission 
following episodes of large amplitude, erratic X-ray variations will be 
found on further occasions, the clue might have been found to relate jet 
formation to X-ray behavior.

\acknowledgments
We thank the unknown referee for a careful reading and detailed comments.
JG is supported by the German Bundesmi\-ni\-sterium f\"ur Bildung, 
Wissenschaft, Forschung und Technologie (BMBF/DARA) under contract No. 
FKZ 50 OR 9201. EHM and RR were supported by NASA contract NAS5-30612.
JG thanks the Deutsche Forschungsgemeinschaft (DFG) for 
a travel grant and is extremely grateful to H. Bradt and R. Vanderspek
for the kind hospitality at MIT where most of this work was done.

\appendix

\newpage

   \begin{figure}
      \caption[liasm]{Light curve of GRS~1915+105 as measured with the RXTE/ASM
           in the 2--12 keV range (top panel) and spectral hardness ratio,
           defined as HR2 = flux (5--12 keV) / flux (3--5 keV) 
           in the lower panel. 
            }
      \label{lcasm}
   \end{figure}

   \begin{figure}
    \caption[lcallobs]{Light curves of GRS~1915+105 as measured with the 
        RXTE/PCA  (2--60 keV range, but dominated by photons below 25 keV
        due to the spectral shape). Shown is the count rate per
        PCUs (note the different scales) at 1 sec time resolution, with each 
        panel showing one orbit of selected pointed observations.
        All the variations in X-ray count rate are due to GRS 1915+105, i.e. 
        Earth occultations and SAA passages are removed, while dead time
        is negligible and has not been corrected for 
        (which is 1\% per 1000 cts/s/PCU).
            }	
      \label{lcall}
   \end{figure}

   \begin{figure}
      \caption[Apr06lull]{Blow up of a part of the light curve of GRS~1915+105 
        on April 6, 1996 showing the quasi-periodically repeating pattern of 
        30 sec duration brightness sputters (top) and the major lull (bottom).
       The top panel of each plot shows the count rate per PCU while
       the lower panel shows the hardness ratio (count ratio in the (4.4-25 keV
       versus 2-4.4 keV band) at the same time resolution of 1 s
       (time along the abscissa refers to the time labeled in the top panel).
       Note the overshooting after the rebound
     and the long waiting time until sputtering resumes.
      Even during the lull, the background rate (about 5 cts/s) is 
    still negligible as compared to the source rate.
            }
      \label{Apr06lull}
   \end{figure}
   \begin{figure}
      \caption[May26lull]{Blow up of a part of the light curve of GRS~1915+105 
        on May 26, 1996.  The top panel of each plot shows the 2--25 keV 
       count rate in one PCU while
       the lower panel shows the hardness ratio (count ratio in the 4.4--25 keV
       versus 2--4.4 keV band) at the same time resolution of 1 s.
	}
      \label{May26lull}
   \end{figure}

\clearpage
\newpage

      \vbox{\psfig{figure=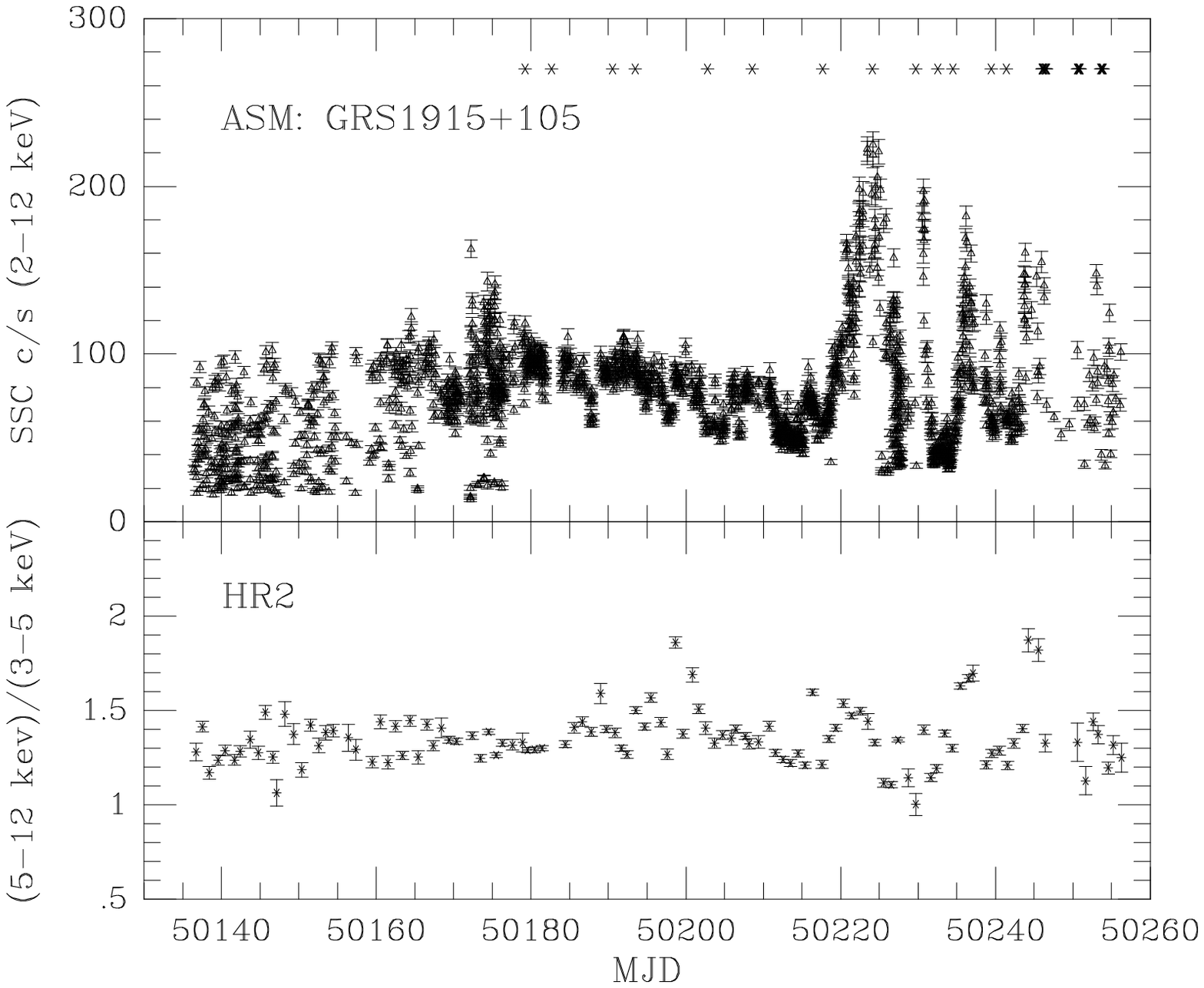,width=14.cm,%
          bbllx=2.3cm,bblly=8.7cm,bburx=18.5cm,bbury=23.9cm,clip=}}\par

\newpage

    \vbox{\psfig{figure=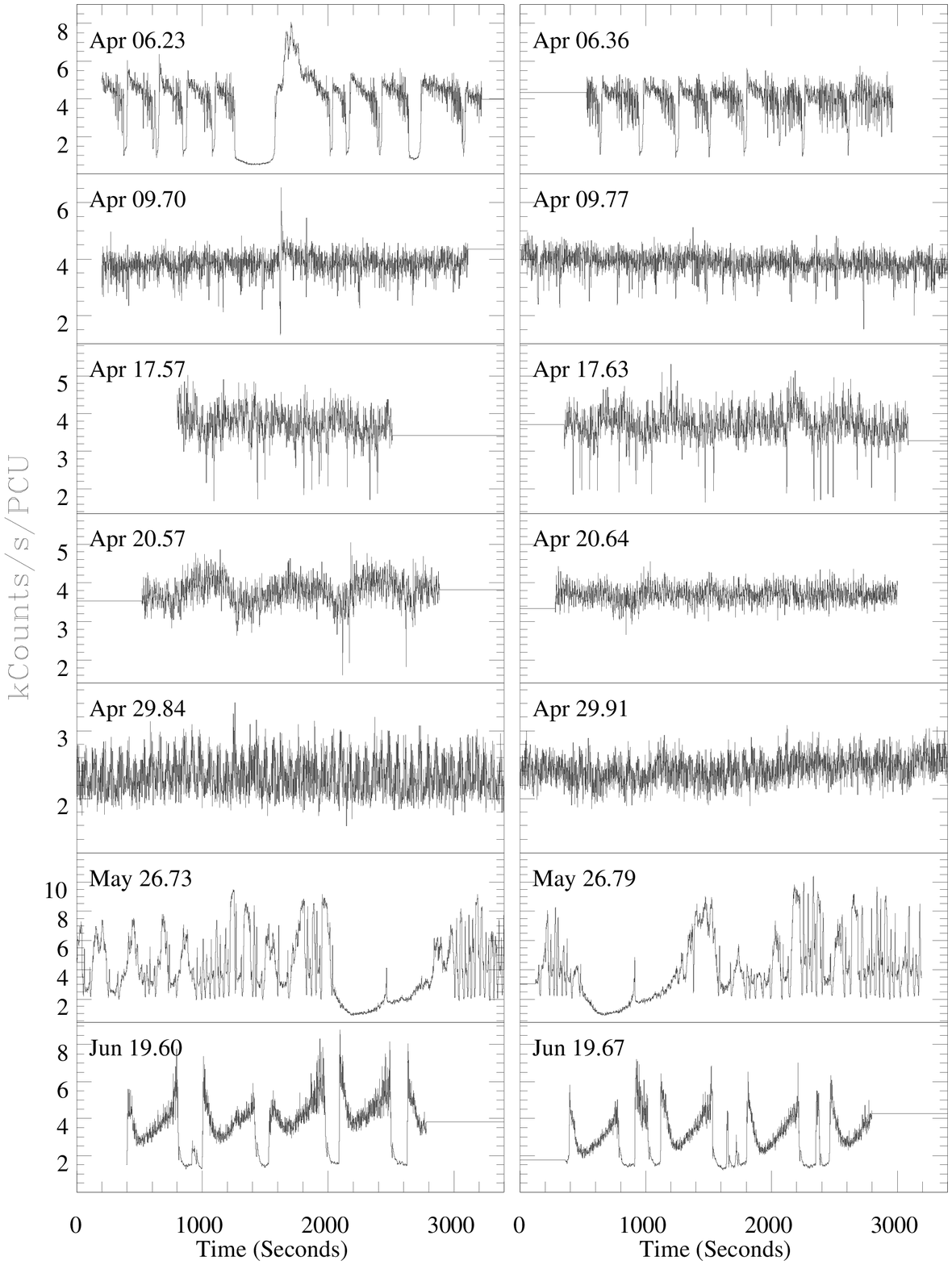,width=14.cm,%
          bbllx=1.4cm,bblly=1.7cm,bburx=19.9cm,bbury=26.4cm,clip=}}\par

\newpage

      \vbox{\psfig{figure=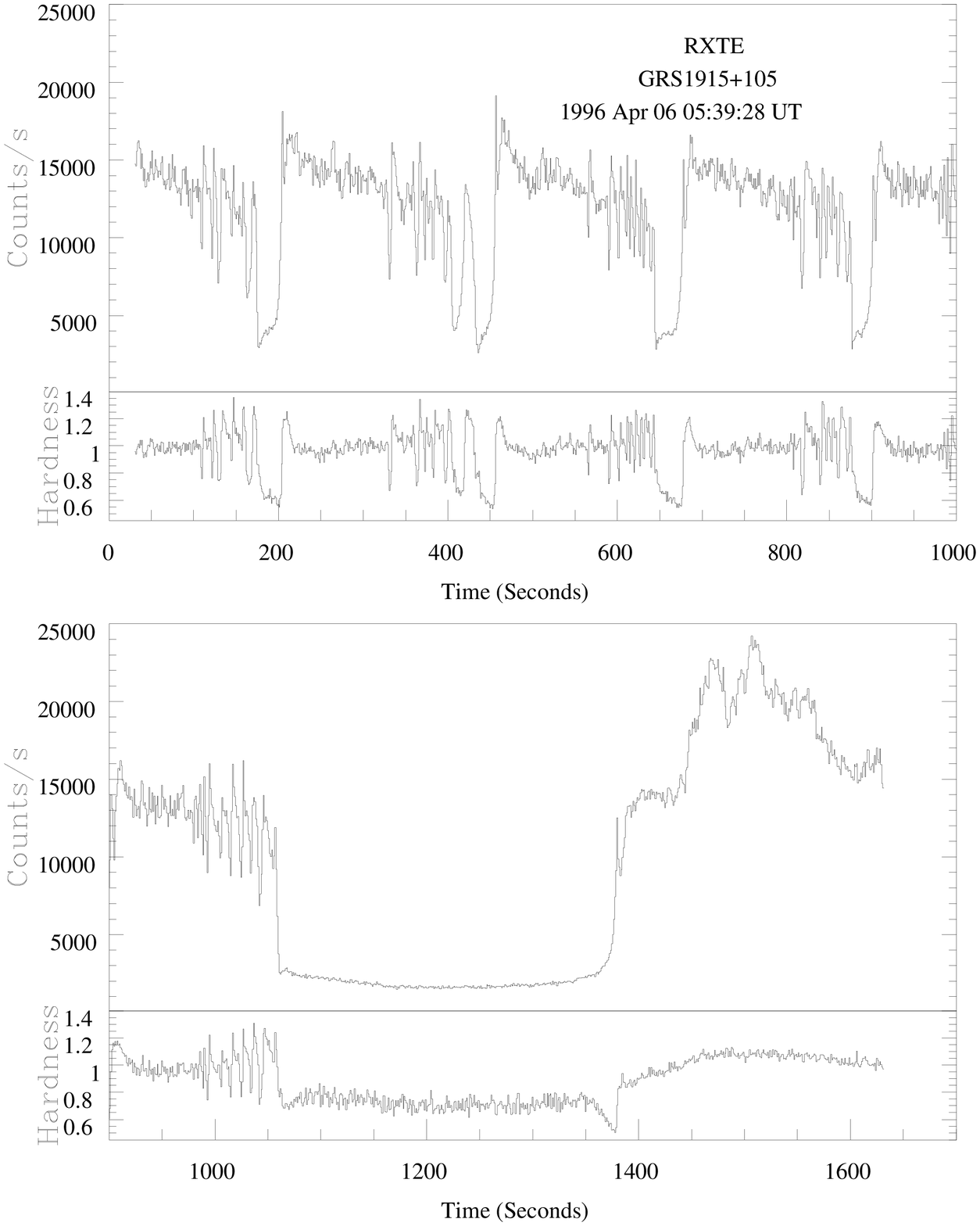,width=12.2cm,%
          bbllx=0.5cm,bblly=1.5cm,bburx=20.3cm,bbury=26.3cm,clip=}}\par

\newpage

      \vbox{\psfig{figure=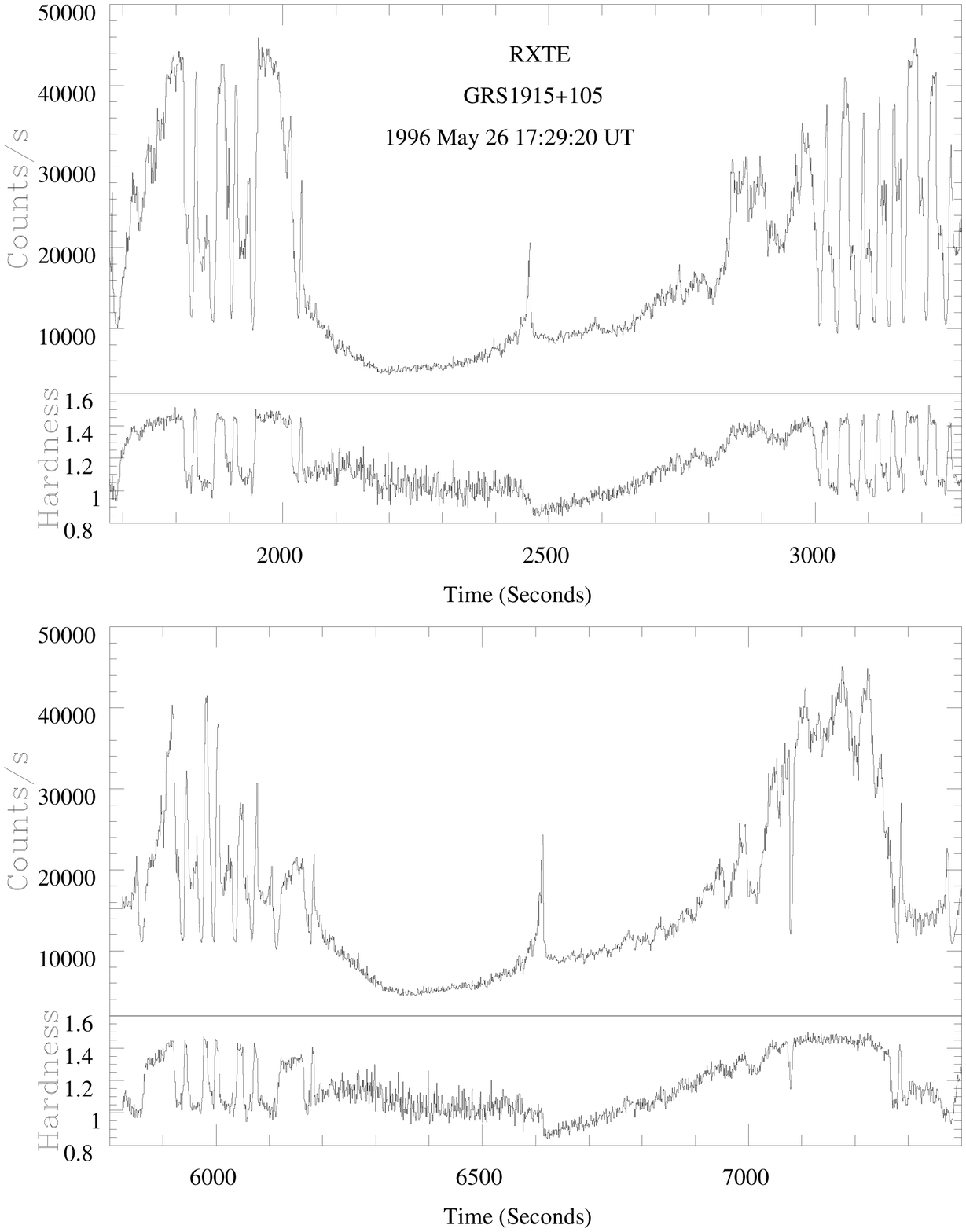,width=12.5cm,%
          bbllx=0.5cm,bblly=1.5cm,bburx=20.cm,bbury=26.3cm,clip=}}\par

\end{document}